# A quantum-operator analysis of N-fold phase accumulation in coherence de Broglie wavelength interferometry

Byoung S. Ham[1,2,*]

[1] Department of Electrical Engineering and Computer Science, Gwangju Institute of Science and Technology, Gwangju, South Korea

[2] Department of Electrical and Computer Engineering, Oregon State University, Corvallis, OR, USA

* Correspondence: bham@gist.ac.kr

## Abstract

Quantum sensing exploits nonclassical states of light, including multiphoton entangled states and squeezed states, to achieve superresolution and/or super-sensitivity. However, its practical implementation is constrained by state-generation complexity, photon loss, and detection efficiency. Here, we present a quantum-operator analysis of coherence de Broglie wavelength (CBW) in an anti-symmetrically coupled Mach-Zehnder interferometer (MZI) architecture and clarify the origin of its superresolution and enhanced phase sensitivity. The two-mode SU(2) structure of the MZI, represented by Pauli operators, shows that successive basis-realigned transformations produce an *N*th-order unitary $U^N(\varphi)$, yielding N-fold phase accumulation $(1 + cosN\varphi)$. The quantum-operator description is particularly significant at the single-photon level, where even a single photon undergoes the *N*th-order unitary evolution through its coherent evolution in the two-mode state space. This differs from entanglement-based quantum sensing based on photonic de Broglie wavelength interferometry, where *N*-fold phase accumulation is associated with an *N*-photon entangled state. The same SU(2) transformation also applies to coherent classical fields, establishing a classical-quantum correspondence of the CBW phase evolution. Fisher-information analysis further shows that the architecture-induced phase transformation yields an order-dependent enhancement of phase sensitivity without increasing the photon number or statistical sampling. CBW therefore provides an architecture-controlled route to high-order interferometric superresolution, accompanied by enhanced phase sensitivity, with potential applications to fringe-counting wavemetry and Fourier-transform-based sensing.

## 1. Introduction

Recently, a fundamentally different approach to sensing and metrology based on the coherence de Broglie wavelength (CBW) has been introduced in the classical regime using a cascaded interferometer architecture [1]. Several proof-of-principle CBW experiments using continuous-wave light [2] and single photons [3] have demonstrated photonic de Broglie wavelength (PBW) [4]-like superresolution. The CBW scheme is based on antisymmetric coupling between identical Mach-Zehnder interferometers (MZIs), each of which obeys SU(2) group symmetry [5]. From the unitary transformation of a single MZI, one might expect that simply cascading identical MZIs would lead to successive phase accumulation. However, direct coupling of two such MZIs can instead result in an identity transformation, thereby preventing the desired phase accumulation for superresolution. CBW overcomes this limitation through successive basis-realigned transformations, yielding the Nth power of the effective MZI unitary operator: $[U(\varphi)]^N = U(N\varphi)$. The quantum-operator description is particularly important at the single-photon level, where the successive unitary transformations demonstrate that even a single photon can undergo *N*-fold phase accumulation through its coherent evolution in the two-mode SU(2) state space. This is fundamentally different from entanglement-based quantum sensing [6-13], particularly N00N-state-based PBW, where *N*-fold phase accumulation is associated with an *N*-photon entangled state. In CBW, by contrast, the order *N* is encoded in the successive interferometric transformations, rather than in the photon number of the input state. The same SU(2) transformation also applies to coherent classical fields, establishing a classical-quantum correspondence of the CBW phase evolution imposed by the linear-optical architecture. Unlike the higher-order photon-coincidence measurements commonly associated with multiphoton entanglement-based quantum sensing [6-13], CBW can be detected through first-order intensity measurements, making it compatible with conventional coherent-light sensing platforms [14-22]. In contrast to the previous coherence-based interpretation [1], here we present a quantum-operator analysis of CBW to clarify how successive basis-realigned transformations generate *N*-fold phase accumulation at the single-photon level without requiring *N*-photon entanglement. The resulting unitary evolution is compared with conventional classical sensing architectures and entanglement-based quantum sensing to clarify the physical origin of CBW superresolution. Furthermore, using classical Fisher information, we derive the CBW-enhanced phase sensitivity and show that the enhancement originates from the architecture-induced phase transformation rather than from an increase of photon number or statistical sampling. The CBW order N therefore represents an interferometric transformation order and should be distinguished from the photon-number resource that determines the conventional shot-noise limit (SNL) [23].

Precision measurements have drawn much attention over decades in both classical [14-22] and quantum sensing regimes [6-13,24-30]. In conventional optical sensing and imaging platforms, spatial resolution is generally governed by the diffraction limit. According to the Rayleigh criterion, two spatially separated point sources are considered just resolvable when the principal maximum of one diffraction pattern coincides with the first minimum of the other [31]. Various nonlinear-optical techniques have been developed to achieve spatial resolution beyond the conventional diffraction limit [14,15]. Such resolution enhancement, however, does not by itself imply a corresponding change in photon-number resource scaling of the phase sensitivity, which generally remains governed by shot-noise statistics for coherent light [23]. In linear-optical interferometric systems, substantially narrower spectral or angular features can also be obtained through coherent multiwave interference, as exemplified by diffraction gratings and optical cavities [31]. In such classical sensing platforms, the resolution enhancement is determined by the number of coherently superposed waves or effective interfering paths and can reach several orders of magnitude. This enhancement is architectural rather than a consequence of statistical resource scaling with the total photon number accumulated over repeated measurement events. Although these linear and nonlinear approaches can produce substantially enhanced resolution, their resolution gain should be distinguished from the statistical precision with which an unknown parameter is estimated. In particular, narrowing an interference or imaging feature does not necessarily provide a proportional improvement in estimation precision when the complete measurement statistics and resources are taken into account. This distinction between resolution and estimation precision is central to the present CBW scheme, although the term "superresolution" is also widely used with a different meaning in high-resolution imaging techniques [14,15]. In practical classical sensors, phase sensitivity is often limited by detector electronic noise. Recent advances in low-noise and single-photon-sensitive detector arrays [32,33] therefore provide a route toward measurements in which electronic noise is sufficiently suppressed for photon-statistical limits.

Quantum sensing generally exploits nonclassical states of light, such as multiphoton entangled states [6-13] and squeezed states [24-30], to enhance measurement performance beyond that achievable with conventional coherent-light interferometry [7]. Squeezed light has been successfully implemented in gravitational-wave detectors such as LIGO, where frequency-dependent squeezing has provided a sensitivity improvement approaching 6 dB [30]. To date, an optical squeezing level of approximately 15 dB has been experimentally demonstrated [24], while optical loss remains a major practical limitation on the observable squeezing level. In parallel, entanglement-based quantum sensing has been extensively studied for achieving PBW-based superresolution [6-13] and sub-shot-noise phase estimation [7-9]. In particular, N00N-state-based PBW interferometry provides an *N*-fold reduction of the effective interference wavelength, $\lambda_{PBW} = \lambda_0/N$, where $\lambda_0$ is the wavelength of the input light. In this case, the *N*-fold phase accumulation originates from an *N*-photon entangled state, resulting in the characteristic PBW superresolution. Its practical scalability, however, is severely constrained by the increasing difficulty of generating, manipulating, and detecting high-order multiphoton entangled states, as exemplified by N00N states demonstrated up to $N = 5$ [33] and GHZ states involving up to 18 photonic qubits [6]. Because such multiphoton states are commonly generated through spontaneous parametric down-conversion (SPDC) and subsequently prepared through linear-optical operations and postselection, their successful generation and detection become increasingly challenging with increasing photon number and entanglement complexity [6,34]. In addition, lower-order harmonic contributions can arise in higher-order correlation measurements, complicating the extraction of the desired *N*-fold interference components [8,9]. Consequently, scaling PBW-based quantum sensing to large *N* remains experimentally challenging [6,33]. Moreover, both squeezed-state and multiphoton-entangled-state sensing are susceptible to optical loss, which degrades their quantum advantage [35,36] and poses a particularly severe challenge for lossy remote-sensing applications such as LiDAR [32]. By contrast, Higgins et al. demonstrated an alternative single-photon approach that does not require multiphoton entanglement, in which multipass interferometry produces an accumulated phase transformation analogous to that considered in the present CBW scheme (see Discussion) [37].

## 2. Results

Figure 1 shows a schematic of CBW based on an anti-symmetrically coupled MZI chain [1]. The light source (L) can be either a conventional laser [2] or a single-photon source [3]. Each MZI constitutes a two-mode system obeying SU(2) group symmetry [5], allowing its unitary transformation to be represented in terms of Pauli operators. For coherent-light illumination, the detected photons can be treated as independent measurement events, providing the conventional statistical-resource framework for Fisher information analysis discussed below [23]. The essential feature of the CBW architecture is the successive coupling of phase-bearing MZIs through dummy MZIs ($\psi = 0$), which realign the exchanged path bases between consecutive interferometers. As analyzed below, this basis realignment causes the individual phase contributions to enter with the same logical sign, producing the $N$th-order unitary evolution $U^N(\varphi)$ and the corresponding $N$-fold phase accumulation. Consequently, an $N$-coupled CBW exhibits the effective wavelength $\lambda_{CBW}^{(N)} = \lambda_0/N$ [1-3]. Importantly, the CBW order $N$ is determined by the successive interferometric transformations rather than by the photon number of the input state. This provides the central distinction from entanglement-based PBW, where the $N$-fold phase dependence originates from an $N$-photon entangled state [4,6-13].

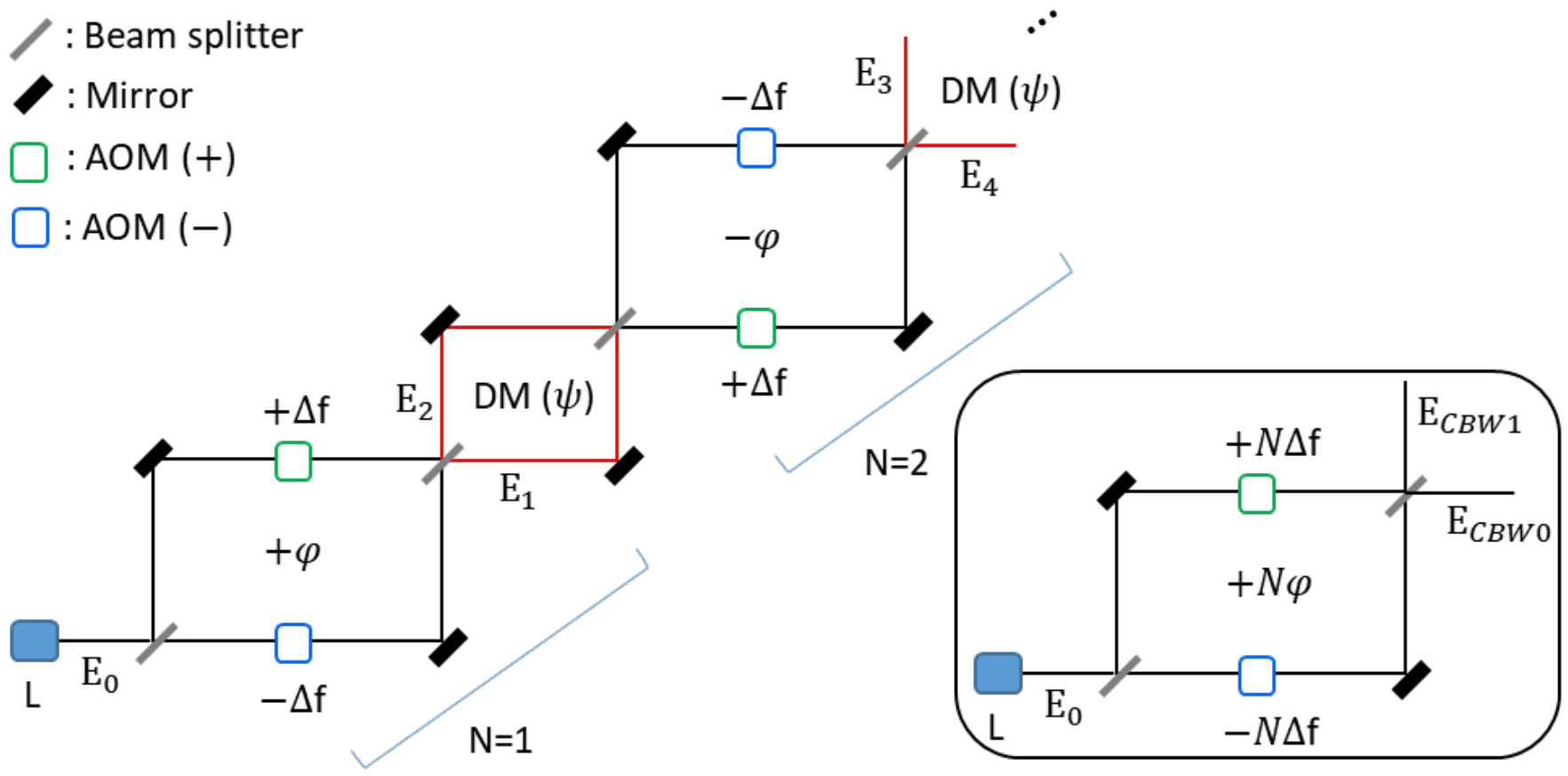

Fig. 1. Schematic of a coherence de Broglie wavelength. DM: dummy MZI, L: laser. Inset: effective MZI. The phase-bearing MZIs are alternatively coupled to DMs with $+\varphi$ and $-\varphi$ phase controls.

For the fully quantum-mechanical analysis of CBW, we introduce the logical basis states corresponding to the two physical paths of the MZI as $|0\rangle = |u\rangle$ and $|1\rangle = |l\rangle$, where $u$ ($l$) denotes the upper (lower) path of the MZI in Fig. 1. These two path modes constitute a two-level system described by SU(2), allowing the MZI transformation to be represented in terms of Pauli operators [5]. In the corresponding Bloch-sphere representation, the Pauli operators $\sigma_z$ and $\sigma_x$ represent the logical path basis and path-basis exchange, respectively. A phase $\varphi$ applied to the upper or lower path is therefore represented by $P_u(\varphi) = \begin{pmatrix} e^{i\varphi} & 0 \\ 0 & 1 \end{pmatrix} = e^{i\varphi/2}e^{i\varphi\sigma_z/2}$ and $P_l(\varphi) = \begin{pmatrix} 1 & 0 \\ 0 & e^{i\varphi} \end{pmatrix} = e^{i\varphi/2}e^{-i\varphi\sigma_z/2}$, respectively. The zero-phase dummy MZI ($\psi = 0$) exchanges the two path bases and, apart from a physically irrelevant global phase, is represented by the Pauli-X operation, $D = M(0) \propto \sigma_x$. Thus, the two alternating subcells of the antisymmetrically coupled MZI structure in Fig. 1 can be represented, after neglecting global phases, as:

$$U_u(\varphi) = \sigma_x e^{i\varphi\sigma_z/2} = cos\frac{\varphi}{2}\sigma_x + sin\frac{\varphi}{2}\sigma_y, \quad (1)$$

$$U_l(\varphi) = \sigma_x e^{-i\varphi\sigma_z/2} = cos\frac{\varphi}{2}\sigma_x - sin\frac{\varphi}{2}\sigma_y. \quad (2)$$

Here, the opposite signs of the $\sigma_y$ terms arise from applying the same phase $\varphi$ to opposite physical paths. On the Bloch sphere, $U_u$ and $U_l$ correspond to two symmetric axes in the x-y equatorial plane, located at azimuthal angles $+\varphi/2$ and $-\varphi/2$, respectively. Thus, $\sigma_y$ does not represent an additional optical operation; rather, it results from combining the $\sigma_z$-generated path phase with the $\sigma_x$ path-basis flip.

The essential role of the dummy MZI follows directly from the Pauli relation:

$$\sigma_x e^{-i\varphi\sigma_z/2}\sigma_x = e^{i\varphi\sigma_z/2}. \tag{3}$$

Equation (3) shows that the path-basis flip converts the nominally opposite relative-phase rotations associated with alternating upper- and lower-path controls into the same logical phase rotation. We define a unit CBW block as one φ-phase-bearing MZI followed by one dummy ($\psi = 0$) MZI, which performs the required path-basis realignment. The second-order CBW therefore consists of two successive unit CBW blocks, with the phase applied alternately to the upper and lower physical paths. Consequently, the second-order CBW becomes:

$$U_{CBW}^{(2)}(\varphi) = DU_l(\varphi)DU_u(\varphi) = e^{i(2\varphi)\sigma_z/2} = [U_u(\varphi)]^2. \tag{4}$$

Thus, although the two phase shifts are physically applied to opposite paths, the intervening dummy-MZI basis realignment causes the two phase contributions to enter with the same logical sign. Equation (4) represents the $N = 2$ CBW and demonstrates explicitly the operator origin of the second-order phase accumulation in Fig. 1.

This result can be generalized directly to an *N*th-order CBW comprising *N* successive unit CBW blocks, each consisting of one phase-bearing MZI and one dummy MZI. With the physical phase applied alternately to the upper and lower paths and the dummy MZI realigning the path basis after each phase-bearing transformation, the successive logical transformations acquire the same phase sign:

$$U_{CBW}^{(N)}(\varphi) = [U_u(\varphi)]^N. \tag{5}$$

Equation (5) identifies the operator origin of the *N*-fold CBW phase accumulation. Here, *N* denotes the number of successive phase-bearing MZI transformations, equivalently the CBW order, rather than the input photon number or the number of statistical measurement events. Each phase-bearing MZI contributes one phase-dependent SU(2) transformation, while the associated dummy MZI performs the path-basis realignment required for successive phase contributions to accumulate with the same logical sign. Consequently, the successive transformations of the unit CBW blocks produce the deterministic *N*-fold phase accumulation. This basis-realignment mechanism is essential to the antisymmetric coupling of Fig. 1 and distinguishes the CBW operator power from a direct cascade of unaligned MZIs.

For a general input path-qubit state of the MZI, $|\Psi_{in}\rangle = \alpha|0\rangle + \beta|1\rangle$, the CBW output state can be represented as $|\Psi_N(\varphi)\rangle = U_{CBW}^{(N)}(\varphi)|\Psi_{in}\rangle$. According to Eq. (5), the successive basis-realigned transformations are equivalent to the *N*th power of the upper-path phase-bearing MZI transformation. In the corresponding phase representation, the CBW state can therefore be written as:

$$\left|U_{CBW}^{(N)}\right\rangle = \frac{1}{\sqrt{2}}\left(e^{iN\varphi}|u\rangle + |l\rangle\right). \tag{6}$$

Equation (6) exhibits the same $N\varphi$ phase dependence as the corresponding nonclassical PBW-based expression, $|\Psi_{PBW}\rangle = (|N,0\rangle + e^{iN\varphi}|0,N\rangle)/\sqrt{2}$, although the physical origins of the two phase responses are fundamentally different. In CBW, the *N*-fold phase dependence results from successive basis-realigned interferometric transformations, whereas in the PBW [4,6-13], it originates from the *N*-photon structure of the nonclassical input state. Thus, CBW realizes *N*-fold phase accumulation without requiring *N*-photon entanglement, analogous to the phase-accumulation mechanism of the single-photon multipass scheme [37].

For the standard CBW input state, $|\Psi_{in}\rangle = |0\rangle$, the output state gives the projection probabilities: $P_0^{(N)}(\varphi) = [1 + \cos(N\varphi)]/2$ and $P_1^{(N)}(\varphi) = [1 - \cos(N\varphi)]/2$. Accordingly, the observable CBW output intensities satisfy $I_0^{(N)} \propto 1 + \cos(N\varphi)$ and $I_1^{(N)} \propto 1 - \cos(N\varphi)$, demonstrating the *N*-fold phase superresolution characteristic of CBW. The fringe period is reduced from $2\pi$ to $2\pi/N$, corresponding to the effective wavelength $\lambda_{CBW} = \lambda_0/N$, where $\lambda_0$ is the wavelength of the input light. Importantly, this superresolution is a direct consequence of the architecture-induced phase accumulation $\varphi \rightarrow N\varphi$, rather than an increase in photon number. The same SU(2) transformation also applies to coherent classical fields because it is governed by the linear-optical evolution of the two path modes. In the absence of photon-photon interactions, each photon undergoes the same unitary transformation independently, consistent with Dirac's description of single-photon interference [38]. The identical transformation of single-photon probability amplitudes and coherent classical-field amplitudes establishes a classical-quantum correspondence of the CBW phase evolution. Thus, the *N*-fold phase accumulation is determined by the basis-realigned interferometric architecture rather than by photon statistics.

The architecture-induced phase multiplication of CBW may also be useful for precision optical metrology. In fringe-counting wavemetry, the *N*-fold phase accumulation increases the number of interference fringes generated for a given optical-path variation, potentially improving wavelength discrimination without increasing the statistical measurement resources [39]. Similarly, in Fourier-transform-based sensing and spectroscopy [40], the increased interferometric phase variation can provide a correspondingly enhanced spectral encoding within a given optical-path-difference range. These features suggest potential applications of CBW to high-resolution wavemetry and Fourier-transform sensing based on architecture-controlled interferometric superresolution.

For the Fisher-information analysis [23], the mean detected signal corresponding to the *N*-fold CBW interference fringe can be represented as:

$$\langle S_N(\varphi)\rangle = \mu[a + bcos(N\varphi)]. \tag{7}$$

where $a$ is a DC offset, $b$ is the modulation amplitude, $\mu$ is the mean detected signal per measurement event. Assuming phase-independent Gaussian signal fluctuations, the Fisher information for $M$ independent and identically distributed (*i.i.d.)* samples is:

$$\begin{aligned} F_N^{(M)}(\varphi) &= \frac{M}{\sigma_S^2}\left(\frac{\partial S_N(\varphi)}{\partial\varphi}\right)^2 \\ &= \frac{M\mu^2 b^2 N^2}{\sigma_S^2} sin^2(N\varphi), \end{aligned} \tag{8}$$

where $\sigma_S^2$ is the signal-noise variance of the fixed input field. At the optimal operating point, $|\sin(N\varphi| = 1$, the Fisher information becomes $F_{N,opt}^{(M)} = M\mu^2 b^2 N^2/\sigma_S^2$. The corresponding Cramer-Rao lower bound (CRLB) is therefore:

$$\Delta\varphi_{CRLB} \geq \frac{1}{\sqrt{F_{N,opt}^{(M)}}} = \frac{\sigma_S}{N\mu|b|\sqrt{M}}. \tag{9}$$

For a fixed input field and a fixed number M of independent measurement events, Eq. (9) exhibits an *N*-fold reduction of the CRLB relative to the single-MZI response under the same signal and noise conditions. Equivalently, at their respective optimal operating points, the architecture-induced phase transformation $\varphi \to N\varphi$ produces the Fisher-information relation $F_{N,opt}^{(M)} = N^2 F_{1,opt}^{(M)}$. The origin of this enhancement, however, must be distinguished from statistical-resource scaling. The CBW order *N* is a fixed architectural parameter specifying the number of coherently coupled phase-bearing MZI transformations, whereas M represents the number of independent measurement events. Fisher information is additive with respect to the latter, satisfying $F_N^{(M)} = MF_N^{(1)}$. For a mean detected photon number $\bar{n}$ per measurement event, the total detected photon resource is $N_{tot} = M\bar{n}$. Accordingly, increasing the statistical photon resource retains the conventional shot-noise scaling, $\Delta\varphi \propto 1/\sqrt{N_{tot}}$, whereas the *N*-dependent factor in Eq. (9) originates from the architecture-induced CBW phase response. The enhanced phase sensitivity of CBW is analogous to that of a conventional high-finesse optical cavity, where coherent multiple reflections produce a multiwave interference effect (see Supplementary Materials) [41]. Importantly, these coherently interfering waves do not constitute *i.i.d.* measurement samples; rather, the sensitivity enhancement arises from the coherent optical architecture itself. Thus, *N* should not be identified with the number of *i.i.d.* samples or with the photon-number resource that determines the conventional SNL. Under the same statistical measurement resource, CBW nevertheless provides an *N*-fold enhancement of local phase sensitivity through the architecture-induced transformation $\varphi \to N\varphi$, analogous to the phase accumulation achieved through polarization-basis realignment in multipass interferometry [37].

# 3. Discussion

The Pauli-operator relations in Eqs. (3)-(5) provide a direct geometric interpretation of CBW phase accumulation on the Bloch sphere of the two-path modes. Taking $|u\rangle$ and $|l\rangle$ as the $+$z and $-$z logical basis states, respectively, the two physical paths constitute a two-level system whose relative phase can be represented by an azimuthal rotation of the corresponding Bloch vector. For a single photon incident on one input port of an MZI, the first balanced beam splitter generates the equal-path coherent superposition: $|\Psi_0\rangle = (|u\rangle + |l\rangle)/\sqrt{2}$, $|u\rangle = \binom{1}{0}$; $|l\rangle = \binom{0}{1}$. The corresponding density matrix is $\rho_0 = |\Psi_0\rangle\langle\Psi_0| = (I + \sigma_x)/2$, and hence the initial-state Bloch vector is $\boldsymbol{r}_0 = (1,0,0)$, lying on the equator along the +x axis. In the first φ-phase-bearing MZI of Fig. 1, the physical phase operator $P_u(\varphi)$ applied to the upper path transforms the single-photon state as $|\Psi_0\rangle \rightarrow \frac{(e^{i\varphi}|u\rangle+|l\rangle)}{\sqrt{2}} \sim \frac{(|u\rangle+e^{-i\varphi}|l\rangle)}{\sqrt{2}}$, where $\sim$ denotes equality up to a physically irrelevant global phase. Accordingly, the Bloch vector representation in the phase-bearing MZI is $\boldsymbol{r}_{MZI} = (cos\varphi, -sin\varphi, 0)$, corresponding to an azimuthal angle $-$φ on the x-y plane of the Bloch sphere. The subsequent dummy MZI exchanges the two path amplitudes, transforming the state into $|\Psi_{DM}\rangle \propto (|u\rangle + e^{i\varphi}|l\rangle)/\sqrt{2}$. Thus, the state vector $\boldsymbol{r}_0$ is changed by the CBW block to $\boldsymbol{r}_{MZI+DM} = (cos\varphi, sin\varphi, 0) \equiv \boldsymbol{r}_{CBW}^{(1)}$. Although the intermediate state within the phase-bearing MZI appears at $-\varphi$ under the adopted logical-basis convention, the combined action of the phase-bearing MZI and the dummy MZI (CBW block) produces a net positive azimuthal rotation $+$φ of the Bloch vector in the equatorial x-y plane.

This effective transformation of one completed CBW block can be represented at the state-vector level by $\left|\Psi_{CBW}^{(1)}\right\rangle = R_z(\varphi)|\Psi_0\rangle$, where $R_z(\varphi) = e^{-i\varphi\sigma_z/2}$, apart from a global phase. Accordingly, $\left|\Psi_{CBW}^{(1)}\right\rangle \sim \frac{(|u\rangle+e^{i\varphi}|l\rangle)}{\sqrt{2}}$ is satisfied. At the Bloch-vector level, the same transformation is represented by $\boldsymbol{r}_{CBW}^{(1)} = \mathcal{R}_z(\varphi)\boldsymbol{r}_0 = (cos\varphi, sin\varphi, 0)$, where $\mathcal{R}_z(\varphi) = \begin{pmatrix} cos\varphi & -sin & 0 \\ sin\varphi & cos\varphi & 0 \\ 0 & 0 & 1 \end{pmatrix}$. Here, $R_z(\varphi)$ is the SU(2) operator acting on the single-photon path state, whereas $\mathcal{R}_z(\varphi)$ is the corresponding SO(3) rotation acting on its Bloch vector. The second CBW block repeats the same basis-realigned transformation, giving $\left|\Psi_{CBW}^{(2)}\right\rangle = [R_z(\varphi)]^2|\Psi_0\rangle = R_z(2\varphi)|\Psi_0\rangle$ and $\boldsymbol{r}_{CBW}^{(2)} = [\mathcal{R}_z(\varphi)]^2\boldsymbol{r}_0 = \mathcal{R}_z(2\varphi)\boldsymbol{r}_0 = (cos2\varphi, sin2\varphi, 0)$. Consequently, N successive CBW transformations give $\left|\Psi_{CBW}^{(N)}\right\rangle = [R_z(\varphi)]^N|\Psi_0\rangle = R_z(N\varphi)|\Psi_0\rangle$, with the corresponding Bloch vector $\boldsymbol{r}_{CBW}^{(N)} = \mathcal{R}_z(N\varphi)\boldsymbol{r}_0 = (cosN\varphi, sinN\varphi, 0)$. Thus, each completed CBW transformation rotates the Bloch vector of the coherent two-path state by $\varphi$ in the equatorial x-y plane, so that *N* successive transformations produce the accumulated rotation of $N\varphi$. This provides the geometric interpretation of Eq. (6), in which the corresponding *N*-fold operator product produces the effective phase coordinate $N\varphi$ and hence the *N*-fold interferometric response shown in the inset of Fig. 1. The CBW phase accumulation can therefore be understood as a path-basis-realignment mechanism that coherently aligns successive phase transformations, rather than as an increase in photon number. This geometric mechanism is analogous to the polarization-basis realignment underlying phase accumulation in multipass interferometry [37], whereas CBW realizes the corresponding SU(2) realignment in the spatial-path basis through the dummy MZIs. In this functional correspondence, the MZI path basis and the dummy MZI in Fig. 1 play roles analogous to the polarization basis and the propagation-reversing mirror, respectively, in ref. 37 [42].

# 3. Conclusions

In conclusion, we presented a quantum-operator analysis of CBW to clarify the physical origin of its high-order interferometric superresolution. The Pauli-operator representation follows from the two-mode SU(2) structure of the MZI rather than from the photon number of the input state. At the single-photon level, the two MZI paths form a two-dimensional path basis, and their coherent superposition provides a direct quantum-state representation of the CBW phase evolution on the Bloch sphere. Successive path-basis-realigned transformations in the coupled interferometric architecture produce an effective rotation, corresponding to an accumulated $N\varphi$ rotation of the two-path state in the equatorial plane. The CBW order *N* is therefore determined by the order of the successive interferometric transformations rather than by the photon number of the input state. The single-photon formulation nevertheless provides an important distinction from PBW superresolution based on entangled states of N00N or GHZ. Whereas the PBW order *N* is encoded in an *N*-photon entangled state, CBW produces an *N*th-order phase response even at the single-photon level through successive basis-realigned transformations. Thus, the role of the single-photon description is not to establish the $N$-fold fringe as an exclusively quantum signature, but to demonstrate that PBW-like superresolution does not require $N$-photon entanglement. The same two-mode SU(2) transformation applies to coherent classical fields, establishing a classical-quantum correspondence and identifying the coherent interferometric transformation itself as the origin of CBW phase multiplication.

The $N$-fold phase multiplication leads to both interferometric superresolution and enhanced phase sensitivity. Fisher-information analysis shows that, at corresponding optimal operating points and under identical signal and noise conditions, the phase information scales with the CBW order as $F_N^{(M)} = N^2 F_1^{(M)}$, resulting in an $N$-fold reduction of the Cramer-Rao lower bound relative to the single-MZI response. This $N$-dependent enhancement originates from the architecture-induced phase transformation, rather than from an increase in the number of independent statistical samples. Accordingly, the CBW order $N$ represents an interferometric transformation order and should be distinguished from the statistical resource governing conventional shot-noise scaling. The $N$-dependent sensitivity enhancement discussed here therefore represents an architecture-induced modulation of the local phase response under fixed statistical measurement conditions, rather than a change in the conventional square-root scaling with statistical resources. Unlike conventional high-resolution mechanisms based on cavity resonance or grating diffraction, CBW directly transforms the interferometric phase coordinate according to $\varphi \to N\varphi$, providing a nonresonant, sinusoidal high-order response. This property is particularly relevant to sensing systems in which information is encoded in fringe number or interferometric frequency. In reference-based wavemetry, the increased fringe density provides a finer phase scale for comparison between reference and probe fields, while in Fourier-transform-based sensing, the transformation $\varphi \to N\varphi$ correspondingly multiplies the interferometric spatial or temporal frequency. The block-sphere representation further shows that this phase multiplication originates from coherent path-basis realignment. This geometric interpretation provides a unified connection between the Pauli-operator description, the observed CBW superresolution, and the architecture-induced enhancement of the phase sensitivity.

# Acknowledgments

The author gratefully acknowledges Prof. Ben Lee at Oregon State University for hosting the author during the sabbatical leave and for kindly providing office space in the Department of Electrical and Computer Engineering.

# Funding

This work was supported by the IITP-ITRC grant (IITP 2026-RS-2021-II211810) funded by the Korean government (Ministry of Science and ICT). BSH also acknowledges that this work was supported by the GIST Research Project grant funded by GIST 2026.

# Conflict of interest

The author is the founder of Qu-Lidar.

# Data availability statement

All data supporting the findings of this publication are available within this article.

# Supplementary materials

The supplementary materials are available at [DOI link].